\documentclass[a4paper]{llncs}
\usepackage[utf8]{inputenc}
\usepackage{default}
\usepackage{graphics}
\usepackage{amssymb}
\usepackage{amsmath}
\usepackage{cleveref}
\usepackage{microtype}
\usepackage{pdfpages}
\usepackage{array}
\usepackage{todonotes}
\usepackage{booktabs}
\usepackage{multirow}
\usepackage{graphicx}
\usepackage{multirow}
\usepackage{url}
\newtheorem{theo}{Theorem}
\newcommand{\OD}{oneDown}
\newcommand{\ND}{noneDown}
\newcommand{\OU}{oneU\!p}
\newcommand{\NU}{noneU\!p}

\newcolumntype{P}{>{\centering\arraybackslash}p{0.37cm}}

\input{drawnets}
\title{New Bounds on Optimal Sorting Networks}
\author{Thorsten Ehlers \and Mike Müller}
\institute{Institut f{\"u}r Informatik, Christian-Albrechts-Universit{\"a}t zu Kiel \\ D-24098 Kiel, Germany.\\
\email{$\{$the,mimu$\}$@informatik.uni-kiel.de}
}

\begin{document}


\maketitle

\begin{abstract}
We present new parallel sorting networks for $17$ to $20$ inputs.
For $17, 19,$ and $20$ inputs these new networks are faster (i.e., they require less computation steps)
than the previously known best networks.
Therefore, we improve upon the known upper bounds for minimal depth sorting networks on $17, 19,$ and $20$ channels.
Furthermore, we show that our sorting network for $17$ inputs is optimal in the sense that no sorting network using less layers exists. 
This solves the main open problem of [D.~Bundala \& J.~Za\'vodn\'y. Optimal sorting networks, Proc. LATA 2014].
\end{abstract}

\section{Introduction}
Comparator networks are hardwired circuits consisting of simple gates that sort their inputs.
If the output of such a network is sorted for all possible inputs, it is called a \emph{sorting network}.
Sorting networks are an old area of interest, and results concerning their size date back at least to the 50's of the last century.

The size of a comparator network in general can be measured by two different quantities:
the total number of comparators involved in the network,
or the number of layers the networks consists of.
In both cases, finding optimal sorting networks (i.e., of minimal size) is a challenging task even when restricted to few inputs, 
which was attacked using different methods.

For instance, Valsalam and Miikkulainen~\cite{ValsalamM13} employed evolutionary algorithms to generate sorting networks with few comparators.
Minimal depth sorting networks for up to $16$ inputs were constructed by Shapiro and Van Voorhis in the 60's and 70's, and by Schwiebert in 2001, who also made use of evolutionary techniques.
For a presentation of these networks see Knuth~\cite[Fig.51]{Knuth}.
However, the optimality of the known networks for $11$ to $16$ channels was only shown recently by Bundala and Z\'avodn\'y~\cite{BundalaZ14},
who partitioned the set of first two layers into equivalence classes and reduced the search to extensions of one representative of each class.
They then expressed the existence of a sorting network with less layers extending these representatives in propositional logic and used a SAT solver to show that the resulting formulae are unsatisfiable.
Codish, Cruz{-}Filipe, and Schneider{-}Kamp~\cite{DBLP:journals/corr/CodishCS14} simplified the generation of the representatives and independently verified Bundala and Z\'avodn\'y's result.

For more than $16$ channels, not much is known about the minimal depths of sorting networks.
Al{-}Haj Baddar and Batcher~\cite{BaddarB09} exhibit a network sorting $18$ inputs using $11$ layers,
which also provides the best known upper bound on the minimal depth of a sorting network for $17$ inputs.
The lowest upper bound on the size of minimal depth sorting networks on $19$ to $22$ channels also stems from a network presented by Al{-}Haj Baddar and Batcher~\cite{BaddarB08}.
For $23$ and more inputs, the best upper bounds to date are established by merging the outputs of smaller sorting networks with Batcher's odd-even merge~\cite{Batcher68}, which needs $\lceil \log n \rceil$ layers for this merging step.

The known lower bounds are due to Parberry \cite{Parberry91} and Bundala and Z\'avodn\'y. 
A new insight by Codish, Cruz{-}Filipe, and Schneider{-}Kamp~\cite{DBLP:journals/corr/CodishCS14a} into the structure of the last layers of sorting networks lead to a significant further reduction of the search space.
Despite all this recently resparked interest in sorting networks, the newly gained insights were insufficient to establish a tight lower bound on the depth of sorting networks for $17$ inputs.

We use the SAT approach by Bundala and Z\'avodn\'y to synthesize new sorting networks of small depths,
and thus provide better upper bounds for $17, 19,$ and $20$ inputs.
Furthermore, our improvements upon their method allow us to raise the lower bound for $17$ inputs. 
Therefore, for the first time after the works of Shapiro, Van Voorhis, and Schwiebert, we present here a new optimal depth sorting network.

An overview of the old and new upper and lower bounds for the minimal depth of sorting networks for up to $20$ inputs is presented in Table~\ref{tbl:bounds}.
\begin{table}[h]
\caption{Bounds on the minimal depth of sorting networks for up to $20$ inputs.}
\label{tbl:bounds}
\centering
\begin{tabular}{r | PPPPPPPPPPPPPPPPPPPP}
\toprule
Number of inputs & 1 & 2 & 3 & 4 & 5 & 6 & 7 & 8 & 9 & 10 & 11 & 12 & 13 & 14 & 15 & 16 & 17 & 18 & 19 & 20 \\
\midrule
Old upper bound & 0 & 1 & 3 & 3 & 5 & 5 & 6 & 6 & 7 & 7 & 8 & 8 & 9 & 9 & 9 & 9 & 11 & 11 & 12 & 12 \\
New upper bound & 0 & 1 & 3 & 3 & 5 & 5 & 6 & 6 & 7 & 7 & 8 & 8 & 9 & 9 & 9 & 9 & \textbf{10} & 11 & \textbf{11} & \textbf{11} \\
Old lower bound & 0 & 1 & 3 & 3 & 5 & 5 & 6 & 6 & 7 & 7 & 8 & 8 & 9 & 9 & 9 & 9 & 9 & 9 & 9 & 9 \\
New lower bound & 0 & 1 & 3 & 3 & 5 & 5 & 6 & 6 & 7 & 7 & 8 & 8 & 9 & 9 & 9 & 9 & \textbf{10} & \textbf{10} & \textbf{10} & \textbf{10} \\
\bottomrule
\end{tabular}
\vspace*{-2em}
\end{table}

\section{Preliminaries}
\begin{figure}[h]
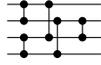

\vspace*{-1em}
    \begin{sortingnetwork}{4}{8}{1}
        \nodeconnection{ {1,2}, {3,4}}
        \addtocounter{sncolumncounter}{2}
        \nodeconnection{ {1,3}}
        \nodeconnection{ {2,4}}
        \addtocounter{sncolumncounter}{2}
        \nodeconnection{{2,3}}
    \end{sortingnetwork}
\vspace*{-2em}
    \caption{A comparator network of depth $3$ with $5$ comparators}
    \label{fig:compnet}
\end{figure}

A \emph{comparator} is a gate with two inputs $in_1$ and $in_2$ and two outputs $out_{\min}$ and $out_{\max}$,
that compares, and if necessary rearranges its inputs such that $out_{\min} = \min\{in_1,in_2\}$ and $out_{\max} = \max\{in_1, in_2\}$.
Combining zero or more comparators in a network yields a \emph{comparator network}.
Comparator networks are usually visualized in a graphical manner in a so-called \emph{Knuth diagram} as depicted in \Cref{fig:compnet}.
Here a comparator connecting two channels is drawn as  
\begin{inlinesortingnetwork}{2}{2}{1}
        \nodeconnection{{1,2}}
    \end{inlinesortingnetwork},
where by convention the upper output is $out_{\min}$ and the lower output is $out_{\max}$. 
A maximal set of comparators with respect to inclusion that can perform parallel comparisons in a comparator network is called a \emph{layer}.
The number of layers of a network is called the \emph{depth} of the network.
A useful tool to verify that a network is a sorting network is the ``0-1-principle'' \cite{Knuth},
which states that a comparator network is a sorting network, if and only if it sorts all binary inputs.

For more details about sorting networks we refer to Knuth~\cite[Sect. 5.3.4]{Knuth}.

\section{Improved Techniques}
In this section we introduce the new techniques and improvements on existing techniques we used to gain our results. 
We will stick to the formulation by Bundala and Z\'avodn\'y~\cite{BundalaZ14}, and introduce new variables if necessary. 
Furthermore, we will extend a technique introduced in their paper, called \emph{subnetwork optimization}. 
It is based on the fact that a sorting network must sort all its inputs, but in order to prove non-existence of sorting networks of a certain depth, 
it is often sufficient to consider only a subset of all possible inputs, which are not all sorted by any network of this restricted depth.
Bundala and Z\'avodn\'y chose subsets of the form $T^{r} = \left\{0^ax1^b \mid |x| = r \text{ and } a+b+|x| = n \right\}$ for $r < n$, which are inputs having a \emph{window} of size $r$.
For an input $0^ax1^b$ from this set the values on the first $a$ channels at any point in the network will always be $0$, and those 
on the last $b$ channels will always be $1$, which significantly reduces the encoding size for these inputs if $a$ and $b$ are sufficiently large. 

\subsection{Prefix optimization}
\begin{figure}[h]
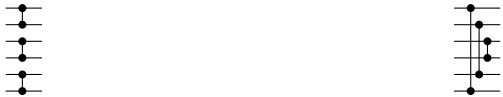

    \begin{minipage}{0.48 \textwidth}
     \begin{sortingnetwork}{6}{2}{1}
        \nodeconnection{ {1,2}, {3,4}, {5,6}}
    \end{sortingnetwork}
    \end{minipage}
    \begin{minipage}{0.48 \textwidth}
     \begin{sortingnetwork}{6}{3}{1}
        \nodeconnection{ {1,6}}
        \nodeconnection{ {2,5}}
        \nodeconnection{ {3,4}}
    \end{sortingnetwork}
    \end{minipage}
    \caption{Pb-style first layer (left), and BZ-style (right) for $6$ channels}
    \label{fig:firstLayer}
\end{figure}

It is a well-known fact that permuting the channels of a sorting network, followed by a repair-procedure called untangling yields another feasible sorting network~\cite{Parberry91}.
In fact, Parberry \cite{Parberry91} used a first layer with comparators of the form $(2i-1, 2i), 1 \leq i \leq \lfloor \frac{n}{2} \rceil$, which we will call \emph{Pb-style} whereas Bundala and Z\'avodn\'y used comparators $(i, n+1-i), 1 \leq i \leq \lfloor \frac{n}{2} \rceil$ in the first layer, which we call \emph{BZ-style}; see \Cref{fig:firstLayer}.
Both versions are equivalent in the sense that if there exists a sorting network $C^d_n$, then there exist sorting networks of the same depth with both of these prefixes. 
Nevertheless, for creating networks obeying a certain prefix as well as proving their non-existence, a given prefix may be hard-coded into the SAT formula. 
Given a prefix $P$ of depth $|P|$, the remaining SAT formula encodes the proposition {\it``There is a comparator network on $n$ channels of depth $d-|P|$ which sorts all outputs of $P$''}. 
Interestingly, the outputs of different prefix styles are not equally handy for the SAT encoding. A Pb-style first layer performs compare-and-swap operations between adjacent channels, 
thus the presorting performed here is more local than the one done by BZ-style first layers. 
Let $out(P)$ denote the set of outputs of a prefix $P$ on $n$ layers. Then, the number of channels that actually must be considered in the SAT formula is given by 
\begin{align*}
 \sum_{ x\in out(P) } \left(n-\max\{a\mid x = 0^ax'\} -\max\{b\mid x = x'1^b\}\right),
\end{align*}
i.e., the sum of window-sizes of all outputs of $P$.

Table~\ref{tab:numRows} shows the impact of these previous deliberations when using a $1$-layer-prefix for $2 \leq n \leq 17$ channels. 
\begin{table}[htb]
\caption{Number of channels to consider in the encoding after the first layer}
\begin{tabular}{l|cccccccccccccccc}
\toprule
n & 2 & 3 & 4 & 5 & 6 & 7 & 8 & 9 & 10 & 11 & 12 & 13 & 14 & 15 & 16 & 17 \\
\midrule
Pb-style & 0 & 5 & 12 & 44 & 84 & 233 & 408 & 1016 & 1704 & 4013 & 6564 & 14948 & 24060 & 53585 & 85296 & 186992 \\
BZ-style & 0 & 4 & 10 & 36 & 72 & 196 & 358 & 876 & 1524 & 3532 & 5962 & 13380 & 22128 & 48628 & 79246 & 171612 \\
\bottomrule
\end{tabular}
\label{tab:numRows}
\end{table}
\begin{table}[htb]
\centering
   \caption{Impact of prefix style when proving that no sorting network for $16$ channels with at most $8$ layers exists.}
   \begin{tabular}{l | r | r}
   \toprule
   Prefix style & Overall time (s) & Maximum time (s) \\
   \midrule
   Pb & 22,241& 326 \\
   BZ & 10,927& 150 \\
   Opt & 5,492 & 36 \\
   \bottomrule
   \end{tabular}
   \label{tab:prefixStyles}
\end{table}

Table ~\ref{tab:prefixStyles} shows running times when proving that no sorting network on $16$ channels with $8$ layers exists. In this case, we used $2$-layer-prefixes according to~\cite{DBLP:journals/corr/CodishCS14}, 
and proved unfeasibility for each of the $211$ distinct prefixes. In the first case, they were permuted and untangled such that the first layer is in Pb-style, whereas the second case
has BZ-style first layers. In the third case, we used an evolutionary algorithm to find a prefix such that the number of channels to consider when using $800$ distinct outputs of the respective prefix
was minimized. As we have $d-|P|$ Boolean variables for each channel which cannot be hard-coded, this procedure minimizes the number of variables in the resulting SAT formula.

This technique reduced the overall running time by factors of $4.05$ and $1.99$, and the maximum running times by $9.0$ and $4.1$, respectively. 
\subsection{Iterative encoding}
\begin{figure}
 \centering
 \includegraphics[scale=0.55]{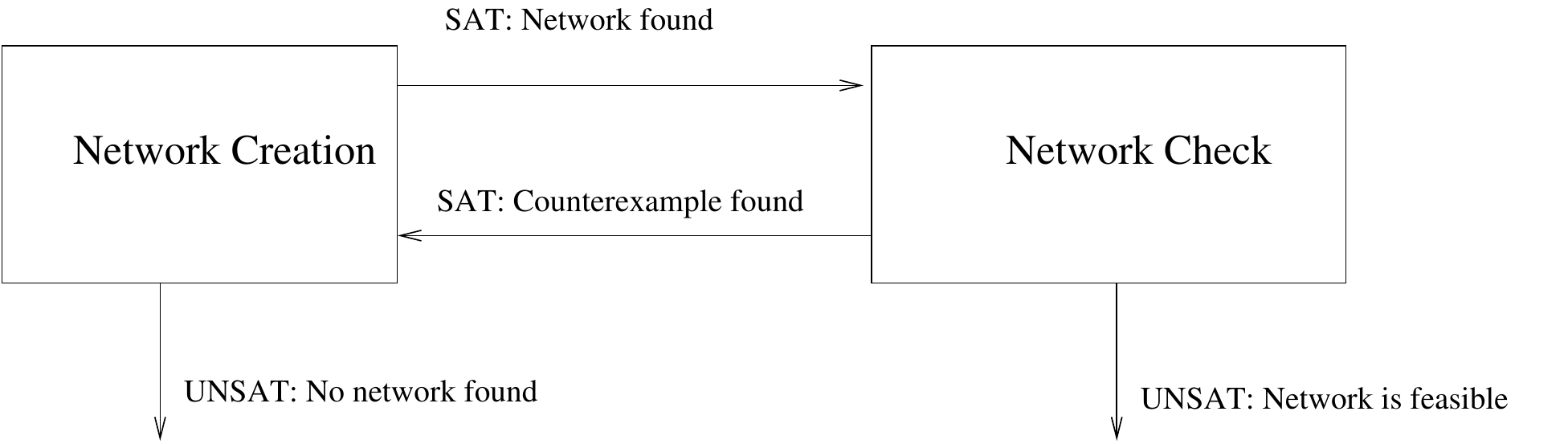}
 \caption{Iterative generation of new inputs}
 \label{fig:nwLoop}
\end{figure}

As mentioned above, it is usually not necessary to use all $2^n$ input vectors to prove lower bounds.
In order to take advantage of this fact, we implemented an iterative approach. 
We start with a formula which describes a feasible comparator network and a (potentially empty) set of initial inputs, and iteratively add inputs until either a feasible sorting network has been found, 
or no network can be found which is able to sort the given set of inputs, as depicted in figure~\ref{fig:nwLoop}.
During the iterative process, counter-examples (i.e., inputs that are not sorted by the network created so far) of minimal window-size are chosen.

Using this technique, we tested the impact of different prefix style on both running time, and the number of inputs required.
\begin{table}[htb]
 \centering
 \caption{Impact of prefix style on running time and number of iterations}
 \begin{tabular}{l|l|rrrrrrrrrr}
 \toprule
  \multicolumn{2}{l|}{Initial Inputs} & 0 & 100 & 200 & 300 & 400 & 500 & 600 & 700 & 800 & 900 \\
  \midrule
  \multirow{2}{*}{Pb} &Time & 157 & 139 & 128 & 86 & 61 & 56 & 45 & 52 & 54 & 59  \\
  & Iterations & 264 & 174 & 72 & 4 & 1 & 1 & 1 & 1 & 1 & 1  \\
  \midrule
  \multirow{2}{*}{BZ} & Time & 88 & 75 & 64 & 47 & 28 & 14 & 14 & 13 & 13 & 19 \\
  &Iterations & 358 & 259 & 165 & 66 & 10 & 1 & 1 & 1 & 1 & 1 \\
  \bottomrule
 \end{tabular}
 \label{tab:testDifferentInputs}
\end{table}
Table~\ref{tab:testDifferentInputs} shows results for one $2$-layer-prefix for $16$ channels, used to prove that this cannot be extended to a sorting network with $8$ layers. 
Here, more inputs are required to prove that a BZ-style prefix cannot be extended when compared to a Pb-style prefix. Nevertheless, BZ-style prefixes are superior in terms of running time. 
Interestingly, the process becomes faster when more inputs than actually required were chosen, this is, using slightly more inputs is beneficial.

Next, we turn to improve the SAT encoding. 
\subsection{Improved SAT encoding}
We modified the SAT encoding of Bundala and Z\'avodn\'y~\cite{BundalaZ14}, significantly reducing the number of clauses. 
A variable $g^k_{i,j}$, with $i < j$, encodes the fact that there is a comparator comparing channels $i$ and $j$ in layer $k$ in the network.
Furthermore, the variable $v^k_i$ stores the value on channel $i$ after layer $k$.
For completeness' sake we list the original encoding completely:
\begin{align*}
 once^k_i(C^d_n) =& \bigwedge_{1 \leq i \neq j \neq l \leq n} \left( \neg g^k_{\min(i, j), \max(i, j) } \vee \neg g^k_{\min(i, l), \max(i, l) } \right) \\
 valid(C^d_n) =& \bigwedge_{1 \leq k \leq d, 1 \leq 1 \leq n} once^k_i(C^d_n) \\
 used^k_i(C^d_n) =& \bigvee_{j <i} g^k_{j, i} \vee \bigvee_{i < j} g^k_{i, j} \\
 update^k_i(C^d_n) =& \left( \neg used_i^k(C^d_n) \rightarrow (v^k_i \leftrightarrow v^{k-1}_i) \right) \wedge \\
      &\bigwedge_{1 \leq j < i} \left( g^k_{j,i} \rightarrow \left(v^k_i \leftrightarrow (v^{k-1}_j \vee v^{k-1}_i )\right) \right) \wedge \\
      &\bigwedge_{i < j \leq n} \left( g^k_{i,j} \rightarrow \left(v^k_i \leftrightarrow (v^{k-1}_j \wedge v^{k-1}_i )\right) \right)
\end{align*}
Here, $once$ encodes the fact that each channel may be used only once in one layer, and $valid$ encodes this constraint for each channel and each layer. 
The $update$-formula describes the impact of comparators on the values stored on each channel after every layer. 
\begin{align*}
 sorts(C^d_n, x) = \bigwedge_{1 \leq i \leq n} (v^0_i \leftrightarrow x_i) \wedge \bigwedge_{1 \leq k \leq d, 1 \leq i \leq n} update^k_i(C^d_n) \wedge \bigwedge_{1 \leq i \leq n} (v^d_i \leftrightarrow y_i) 
\end{align*}
The constraint $sorts$ encodes if a certain input is sorted by the network $C^d_n$.
For this purpose, the values after layer $d$ (i.e., the outputs of the network) are compared to the vector $y$, which is a sorted copy of the input $x$.
A sorting network for $n$ channels on $d$ layers exists iff $valid(C^d_n) \wedge \bigwedge_{x \in \{0, 1\}^n} sorts(C^d_n, x)$ is satisfiable. 

Consider an input $x = 0^ax'1^b$, and a comparator $g^k_{i, j}$ with $i \leq a$. This is, we have $v^k_i \leftrightarrow 0 \wedge v^{k-1}_j \equiv 0$, and 
$v^k_j \leftrightarrow 0 \vee v^{k-1}_j \equiv v^{k-1}_j$. As the same holds for $j > n-b$, we have that comparators ''leaving`` a subnetwork need not be considered 
for sorting the respective inputs. Furthermore, if $v^{k-1}_i \leftrightarrow 1$ for some $k$ and $i$, using any comparator $g^k_{j, i}$ will cause $v^k_i \leftrightarrow 1$. 
Thus, for every channel $i$ we introduce $\OD^k_{i, j}$- and $\OU^k_{i,j}$-variables which indicate whether there is a comparator $g^k_{l,j}$ for some $i \leq l < j$ or $g^k_{i, l}$ for some $i < l \leq j$, respectively. 
\begin{align*}
    \OD^k_{i, j} &\leftrightarrow \bigvee_{i < l \leq j} g^k_{i, l} & \ND^k_{i, j} &\leftrightarrow \neg \OD^k_{i, j} \\
    \OU^k_{i, j} &\leftrightarrow \bigvee_{i \leq l < j} g^k_{l, j} & \NU^k_{i, j} &\leftrightarrow \neg \OU^k_{i, j}
\end{align*}
To make use of these new predicates, given an input $x = 0^ax'1^b$, for all $a < i \leq n-b$ we add
\begin{align*}
 v^{k-1}_i \wedge \ND^k_{i, n-b} &\rightarrow v^{k}_i \\
 \neg v^{k-1}_i \wedge \NU^k_{a+1, i} &\rightarrow \neg v^{k}_i,  
\end{align*}
to the formula and remove all update-constraints that are covered by these constraints. This offers several advantages: Firstly, it reduces the size of the resulting formula in terms of both the number of clauses, 
and the overall number of literals. Secondly, this encoding allows for more propagations, thus, conflicts can be found earlier. Thirdly, it offers a more general perspective on 
the reasons of a conflict. Table~\ref{tab:allResultsFor16} shows the impact of both the new encoding and permuting the prefix, which results in an average speed-up of $8.2$, and a speed-up of $17.1$ for the hardest 
prefixes. 
\begin{table}
\caption{Results for different settings when proving the non-existence of $8$-layer sorting-networks for $16$ channels}
\label{tab:allResultsFor16}
\centering
 \begin{tabular}{l|l|r|r|r|r|r}
 \toprule
  Encoding & Prefix-Style & Overall time (s) & Max. time (s) & Variables & Clauses & Literals \\
  \midrule
  \multirow{3}{*}{Old} & Pb & 22,241 & 326 & 108,802 & 4,467,201 & 13,977,393 \\
   & BZ & 10,927 & 150 & 99,850 & 3,996,902 & 12,442,522 \\
   & Opt & 5,492 & 36 & 84,028 & 3,183,363 & 9,879,588 \\
   \midrule
   \multirow{3}{*}{New} & Pb & 11,766 & 196 & 110,398 & 2,443,186 & 8,108,501 \\
   & BZ & 4,359 & 54 & 101,404 & 2,049,744 & 6,799,486 \\
   & Opt &  2,702 & 19 & 85,652 & 1,504,177 & 4,981,882 \\
   \bottomrule
 \end{tabular}

\end{table}

\section{Obtaining new lower bounds}
In a first test, we tried to prove that there is no sorting network on $17$ channels using at most $9$ layers by using a SAT encoding which 
was almost identical to the one introduced in~\cite{BundalaZ14}, enriched by constraints on the last layers~\cite{DBLP:journals/corr/CodishCS14a}.
Before we broke up this experiment after $48$ days, we were able to prove that $381$ out of $609$ prefixes cannot be extended to sorting networks of depth $9$.
Showing unsatisfiability of these formulae took $353 \cdot 10^6$ seconds of CPU time, with a maximum of $3 \cdot 10^6$ seconds. 

In a new attempt, we used a modified encoding as described in the previous section. For every equivalence class of $2$-layer-prefixes, we chose a representative which minimizes the number of 
variables in the SAT encoding when using $2,000$ distinct inputs. This time, we were able to prove that none of the prefixes can be extend to a sorting network using $9$ layers. 
The overall CPU time for all $609$ equivalence classes was $27.63 \cdot 10^6$ seconds with a maximum running time of $97,112$ seconds.
This is a speed up of at least $42.7$ concerning the maximum running time, and $20.4$ for the average running time. 
Since the result for all $2$-layer-prefixes was unsat, we conclude: 
\begin{theo}
 Any sorting network for $n \geq 17$ channels has at least $10$ layers.
\end{theo}

\section{Finding faster networks}
Even though SAT encodings for sorting networks as well as SAT solvers themselves have become much better within the last years, 
generating new, large sorting networks from scratch is still out of their scope. Hence, we extended ideas by Al-Haj Baddar and Batcher~\cite{BaddarB09}. 
\subsection{Using hand-crafted prefixes}
\begin{figure}[h]
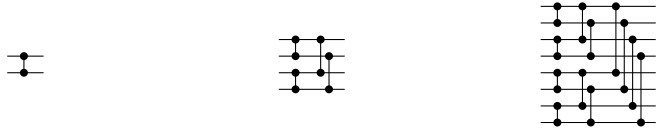

    \begin{minipage}{0.3 \textwidth}
     \begin{sortingnetwork}{2}{2}{1}
        \nodeconnection{ {1,2}}
    \end{sortingnetwork}
    \end{minipage}
   \begin{minipage}{0.3 \textwidth}
     \begin{sortingnetwork}{4}{5}{1}
        \nodeconnection{ {1,2}, {3,4}}
        \addtocounter{sncolumncounter}{2}
        \nodeconnection{ {1,3}}
        \nodeconnection{ {2,4}}
    \end{sortingnetwork}
    \end{minipage}
    \begin{minipage}{0.3 \textwidth}
     \begin{sortingnetwork}{8}{10}{1}
        \nodeconnection{ {1,2}, {3,4}, {5, 6}, {7, 8}}
        \addtocounter{sncolumncounter}{2}
        \nodeconnection{ {1,3}, {5, 7}}
        \nodeconnection{ {2,4}, {6, 8}}
        \addtocounter{sncolumncounter}{2}
        \nodeconnection{ {1,5}}
        \nodeconnection{ {2,6}}
        \nodeconnection{ {3,7}}
        \nodeconnection{ {4,8}}
    \end{sortingnetwork}
    \end{minipage}
    
    \caption{Generating partially ordered sets for $n \in \{2,4,8\}$ inputs.}
    \label{fig:posets}
\end{figure}

A well-known technique for the creation of sorting networks is the generation of partially ordered sets for parts of the input in 
the first layers. 
\Cref{fig:posets} shows comparator networks which create partially ordered sets for $2$, $4$ and $8$ input bits.
In the case of $n=2$, the output will always be sorted. For $n=4$ bits, the set of possible output vectors is given by %
$$\left\{\begin{pmatrix} 0 & 0 & 0 & 0 \end{pmatrix}^T,
\begin{pmatrix} 0 & 0 & 0 & 1 \end{pmatrix}^T,
\begin{pmatrix} 0 & 0 & 1 & 1 \end{pmatrix}^T,
\begin{pmatrix} 0 & 1 & 0 & 1 \end{pmatrix}^T,
\begin{pmatrix} 0 & 1 & 1 & 1 \end{pmatrix}^T,
\begin{pmatrix} 1 & 1 & 1 & 1 \end{pmatrix}^T \right\},
$$ i.e., there are $6$ possible outputs. Furthermore, the first output bit will equal zero unless all input bits are set to one, 
and the last output bit will always be set to one unless all input bits equal zero. Similarly, the network for $n=8$ inputs allows for $20$ different output vectors. 
Prefixes of sorting networks which consist of such snippets are referred to as \emph{Green filters}~\cite{DBLP:conf/gecco/Coles12}. 


%
%
%
%
%


\subsection{Results}
\begin{figure}[h]
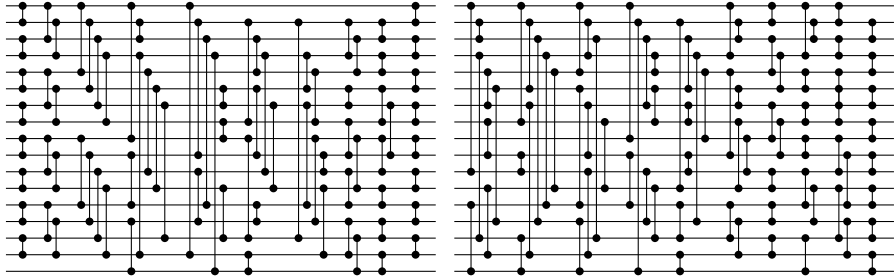

    \begin{minipage}{.48\textwidth}
    \begin{sortingnetwork}{17}{42}{1}
        \nodeconnection{ {1,2}, {3,4}, {5,6}, {7,8}, {9,10}, {11,12}, {13,14}, {15,16}}
        \addtocounter{sncolumncounter}{2}
        \nodeconnection{ {1,3}, {5,7}, {9,11}, {13,15}}
        \nodeconnection{ {2,4}, {6,8}, {10,12}, {14,16}}
        \addtocounter{sncolumncounter}{2}
        \nodeconnection{ {1,5}, {9,13}}
        \nodeconnection{ {2,6}, {10,14}}
        \nodeconnection{ {3,7}, {11,15}}
        \nodeconnection{ {4,8}, {12,16}}
        \addtocounter{sncolumncounter}{2}
        \nodeconnection{ {1,9}, {10,13}, {14,17}}
        \nodeconnection{ {2,3}, {4,16}}
        \nodeconnection{ {5,11}}
        \nodeconnection{ {6,12}}
        \nodeconnection{ {7,15}}
        \addtocounter{sncolumncounter}{2}
        \nodeconnection{ {1,16}}
        \nodeconnection{ {2,10}, {11,14}}
        \nodeconnection{ {3,13}}
        \nodeconnection{ {4,17}}
        \nodeconnection{ {6,7}, {8,9}, {12,15}}
        \addtocounter{sncolumncounter}{2}
        \nodeconnection{ {2,8}, {9,15}, {16,17}}
        \nodeconnection{ {3,5}, {6,10}, {13,14}}
        \nodeconnection{ {4,11}}
        \nodeconnection{ {7,12}}
        \addtocounter{sncolumncounter}{2}
        \nodeconnection{ {2,15}}
        \nodeconnection{ {4,6}, {7,13}}
        \nodeconnection{ {5,8}, {9,14}}
        \nodeconnection{ {10,11}, {12,16}}
        \addtocounter{sncolumncounter}{2}
        \nodeconnection{ {2,4}, {6,7}, {8,10}, {11,13}, {14,16}}
        \nodeconnection{ {3,5}, {9,12}, {15,17}}
        \addtocounter{sncolumncounter}{2}
        \nodeconnection{ {2,3}, {4,5}, {6,8}, {9,11}, {12,13}, {14,15}, {16,17}}
        \nodeconnection{ {7,10}}
        \addtocounter{sncolumncounter}{2}
        \nodeconnection{ {1,2}, {3,4}, {5,6}, {7,8}, {9,10}, {11,12}, {13,14}, {15,16}}
        
    \end{sortingnetwork}
    \end{minipage}
    \begin{minipage}{.48\textwidth}
            \begin{sortingnetwork}{17}{43}{1}
       \nodeconnection{{ 1,11}, {13,17}}
\nodeconnection{{ 2,3}, { 4,15}}
\nodeconnection{{ 5,7}, { 8,10}, { 12,16}}
\nodeconnection{{ 6,14}}
\addtocounter{sncolumncounter}{2}
\nodeconnection{{ 1,8}, { 10,11},{ 15,17}}
\nodeconnection{{ 2,6}, { 7,16}}
\nodeconnection{{ 3,14}}
\nodeconnection{{ 4,13}}
\nodeconnection{{ 5,12}}
\addtocounter{sncolumncounter}{2}
\nodeconnection{{ 1,5}, { 6,13}, { 14,17}}
\nodeconnection{{ 2,4}, { 7,10}, { 11,16}}
\nodeconnection{{ 3,15}}
\nodeconnection{{ 8,12}}
\addtocounter{sncolumncounter}{2}
\nodeconnection{{ 1,9}, { 10,13}}
\nodeconnection{{ 2,17}}
\nodeconnection{{ 3,7}, { 11,14}}
\nodeconnection{{ 4,5}, { 6,8}, { 12,15}}
\addtocounter{sncolumncounter}{2}
\nodeconnection{{ 2,12}, { 13,15}, { 16,17}}
\nodeconnection{{ 3,6}, { 7,8}, { 10,11}}
\nodeconnection{{ 4,14}}
\nodeconnection{{ 5,9}}
\addtocounter{sncolumncounter}{2}
\nodeconnection{{ 1,4}, { 5,10}, { 13,15}}
\nodeconnection{{ 2,3}, { 6,7}, { 8,12}, { 14,16}}
\nodeconnection{{ 9,11}}
\addtocounter{sncolumncounter}{2}
\nodeconnection{{ 1,2}, { 3,5}, { 7,9}, { 11,12}, { 13,14}, { 15,16}}
\nodeconnection{{ 4,6}, { 8,10}}
\addtocounter{sncolumncounter}{2}
\nodeconnection{{ 1,4}, { 5,6}, { 7,8}, { 9,10}, { 11,13}, { 15,17}}
\nodeconnection{{ 2,3}, { 12,14}}
\addtocounter{sncolumncounter}{2}
\nodeconnection{{ 1,2}, { 3,4}, { 5,7}, { 6,8}, { 9,11}, { 12,15}}
\nodeconnection{{ 10,13}, { 14,16}}
\addtocounter{sncolumncounter}{2}
\nodeconnection{{ 2,3}, { 4,5}, { 6,7}, { 8,9}, { 10,11}, { 12,13}, { 14,15}, { 16,17}}
\addtocounter{sncolumncounter}{2}

    \end{sortingnetwork}
\end{minipage}
    \caption{Sorting networks for $17$ channels of depth $10$.}
    \label{fig:17_10}
\end{figure}

We present two sorting networks improving upon the known upper bounds on the minimal depth of sorting networks.
The networks presented in \Cref{fig:17_10} are sorting networks for $17$ channels using only $10$ layers,
which outperform the currently best known network due to Al{-}Haj Baddar and Batcher~\cite{BaddarB09}.
The first three layers of the network on the left are a Green filter on the first $16$ channels, the remainder of this network was created by a SAT solver.
To create the network on the right, we applied the prefix optimization procedure described earlier to the Green filter prefix. 
The first version of our solver required $29921$ seconds to find this network, whereas our current solver can find these networks in $282$ seconds, when using the Green filter, and $60$ seconds when using the optimized prefix.

Thus, we can summarize our results in the following theorem:
\begin{theo}
 The optimum depth for a sorting network on $17$ channels is $10$.
\end{theo}

\begin{figure}
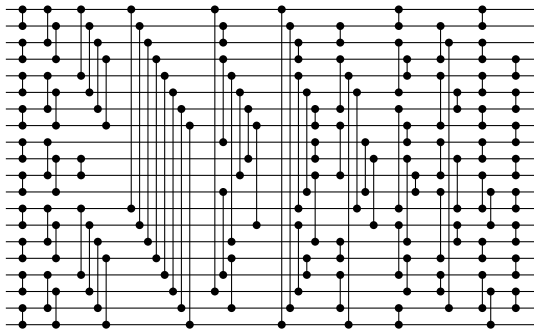

    \begin{sortingnetwork}{20}{52}{1}
        \nodeconnection{ {1,2}, {3,4}, {5,6}, {7,8}, {9,10}, {11,12}, {13,14}, {15,16}, {17,18}, {19,20}}
        \addtocounter{sncolumncounter}{2}
        \nodeconnection{ {1,3}, {5,7}, {9,11}, {13,15}, {17,19}}
        \nodeconnection{ {2,4}, {6,8}, {10,12}, {14,16}, {18,20}}
        \addtocounter{sncolumncounter}{2}
        \nodeconnection{ {1,5}, {10,11}, {13,17}}
        \nodeconnection{ {2,6}, {14,18}}
        \nodeconnection{ {3,7}, {15,19}}
        \nodeconnection{ {4,8}, {16,20}}
        \addtocounter{sncolumncounter}{2}
        \nodeconnection{ {1,13}}
        \nodeconnection{ {2,14}}
        \nodeconnection{ {3,15}}
        \nodeconnection{ {4,16}}
        \nodeconnection{ {5,17}}
        \nodeconnection{ {6,18}}
        \nodeconnection{ {7,19}}
        \nodeconnection{ {8,20}}
        \addtocounter{sncolumncounter}{2}
        \nodeconnection{ {1,18}}
        \nodeconnection{ {2,3}, {4,9}, {12,17}}
        \nodeconnection{ {5,15}, {16,19}}
        \nodeconnection{ {6,11}}
        \nodeconnection{ {7,10}}
        \nodeconnection{ {8,14}}
        \addtocounter{sncolumncounter}{2}
        \nodeconnection{ {1,20}}
        \nodeconnection{ {2,19}}
        \nodeconnection{ {3,4}, {5,13}, {14,18}}
        \nodeconnection{ {6,12}, {16,17}}
        \nodeconnection{ {7,8}, {9,10}, {11,15}}
        \addtocounter{sncolumncounter}{2}
        \nodeconnection{ {2,3}, {4,7}, {8,11}, {15,16}, {17,19}}
        \nodeconnection{ {5,20}}
        \nodeconnection{ {6,13}}
        \nodeconnection{ {9,12}}
        \nodeconnection{ {10,14}}
        \addtocounter{sncolumncounter}{2}
        \nodeconnection{ {1,2}, {3,6}, {7,13}, {14,17}, {19,20}}
        \nodeconnection{ {4,5}, {8,9}, {10,15}, {16,18}}
        \nodeconnection{ {11,12}}
        \addtocounter{sncolumncounter}{2}
        \nodeconnection{ {2,4}, {5,8}, {9,11}, {12,16}, {17,18}}
        \nodeconnection{ {3,19}}
        \nodeconnection{ {6,7}, {10,13}, {14,15}}
        \addtocounter{sncolumncounter}{2}
        \nodeconnection{ {1,2}, {3,4}, {5,6}, {7,8}, {9,10}, {11,13}, {15,16}, {17,19}}
        \nodeconnection{ {12,14}, {18,20}}
        \addtocounter{sncolumncounter}{2}
        \nodeconnection{ {4,5}, {6,7}, {8,9}, {10,11}, {12,13}, {14,15}, {16,17}, {18,19}}
        
    \end{sortingnetwork}
    \vspace*{-5mm}
    \caption{A sorting network for $20$ channels of depth $11$.}
    \label{fig:20_11}
\end{figure}

The network displayed in \Cref{fig:20_11} sorts $20$ inputs in $11$ parallel steps, 
which beats the previously fastest network using $12$ layers~\cite{BaddarB08}.
In the first layer, partially ordered sets of size $2$ are created. These are merged to $5$ partially ordered sets of size $4$ in the second layer. 
The third layer is used to create partially ordered sets of size $8$ for the lowest and highest wires, respectively. These are merged in the fourth layer.

The wires in the middle of the network are connected in order to totally sort their intermediate output. 
Using this prefix and the necessary conditions on sorting networks depicted above, we were able to create the remaining layers using our iterative, SAT-based approach. 
Interestingly, the result was created in $588$ iterations, thus $587$ different input vectors were sufficient.

\section{Tools}
Our software is based on the well-known SAT solver MiniSAT 2.20. 
Before starting a new loop of our network creation process, we used some probing-based preprocessing techniques~\cite{DBLP:conf/ictai/LynceS03} 
as they were quite successful on this kind of SAT formulae.
MiniSAT uses activity values for clauses which are used for managing the learnt clause database. Here, we changed the value ``cla-decay'' to $0.9999$, 
which leads to better control on learnt clauses that were not used for a long time.
Our experiments were performed on Intel Xeon E5-4640 CPUs clocked at 2.40GHz.
The software used for our experiments can be downloaded at \url{http://www.informatik.uni-kiel.de/~the/SortingNetworks.html}.

\section*{Acknowledgments}
We would like to thank Michael Codish, Lu{\'{\i}}s Cruz{-}Filipe, and Peter Schneider{-}Kamp for fruitful discussions on the subject and their valuable comments on an earlier draft of this article.
Furthermore, we thank Dirk Nowotka for providing us with the computational resources to run our experiments.

\bibliographystyle{abbrv}
\bibliography{bib}
\end{document}